\documentclass[11pt,twoside]{article}

\usepackage{asp2006}
\usepackage{epsf}
\usepackage{psfig}
\usepackage{lscape}
\usepackage{graphicx} 

\begin{document}
\title{Stellar Populations in Field Early--Type Galaxies}   
\author{F. Annibali}
\affil{STSCI, 3700 San Martin Drive, Baltimore, MD 21218, USA}
\author{A. Bressan, R. Rampazzo}
\affil{INAF - Osservatorio Astronomico di Padova, Vicolo dell'Osservatorio 5, 35122 Padova, Italy}
\author{W. Zeilinger}
\affil{Institut f\" ur Astronomie der Universit\" at  Wien, T\"urkenschanzstra$\ss$e 17, A-1180 Wien, Austria }   
\author{L. Danese}
\affil{SISSA, Via Beirut 4, 34014 Trieste, Italy}

\begin{abstract} 
We have acquired intermediate resolution spectra in the 3700-7000 \AA
wavelength range for a sample of 65 early-type galaxies
predominantly located in low density environments, a
large fraction of which show emission lines. The spectral coverage
and the high quality of the spectra allowed us to derive Lick
line-strength indices and to study their behavior at different
galacto-centric distances. Ages, metallicities and element abundance
ratios have been derived for the galaxy sample by comparison of the
line-strength index data set with our new developed Simple Stellar
Population (SSP) models. We have analyzed the behavior of the derived
stellar population parameters with the central galaxy velocity
dispersion and the local galaxy density in order to understand
the role played by mass and environment on the evolution of early-type
galaxies. We find that the chemical path is mainly driven by
the halo mass, more massive galaxies exhibiting the more efficient
chemical enrichment and shorter star formation timescales. Galaxies in
denser environments are on average older than galaxies in less dense
environments. The last ones show a large age spread which is likely
to be due to rejuvenation episodes. 
\end{abstract}



\vspace{-0.5cm}

\section{Introduction}

More than half of all stars in the Local Universe are found in massive 
spheroidal components (e.g. \citet{fug98}).
To understand the history of assembling of the bulk of stellar mass in the
Universe is thus fundamental to derive how and when the present day 
massive early--type galaxies built up. 
However, the formation and evolution of early--type galaxies is still a 
matter of debate. 
According to the standard $\Lambda$ Cold Dark Matter scenario ($\Lambda$ CDM)
for structure formation, galaxies built up their present day mass through a continuous assembly driven by the hierarchical merging of dark matter halos 
(e.g.\citet{wr78}).
In this scenario, massive early-type galaxies appear 
rather late in the  history
of the Universe as the culmination of a hierarchical merging process.
On the other hand, in the traditional monolithic collapse scenario 
(e.g. \cite{egg62}; \cite{tins72})
the spheroidal component forms by the gravitational collapse 
of a gas cloud at relatively high redshifts.
As a result of the rapidity of this collapse,
the bulk of stars in ellipticals should be relatively old.
The study of absorption line indices in local 
early--type galaxies has proven to be one of the most
powerful diagnostics to constrain star formation history
and to trace star evolution over time. Besides "direct" age estimates,
indirect evidences of the duration of the star formation
process can be analyzed through the study of
the chemical enrichment pattern, and more specifically of
the $\alpha$/Fe enhancement.

Here we present a study of the stellar populations in a sample 
65 nearby early-type galaxies, 
predominantly located in low density environments.
We look in particular for correlations of the stellar population parameters
with galaxy mass and environment, in order to understand the role played by
them on the evolution of massive early-type systems.
This work has been presented in three separate papers
(\cite{ram05}, Paper I; \cite{anni06a,anni06b}, Papers~II and III) 
to which we refer for details.

\vspace{-0.5cm}

 \begin{figure}
 \plotone{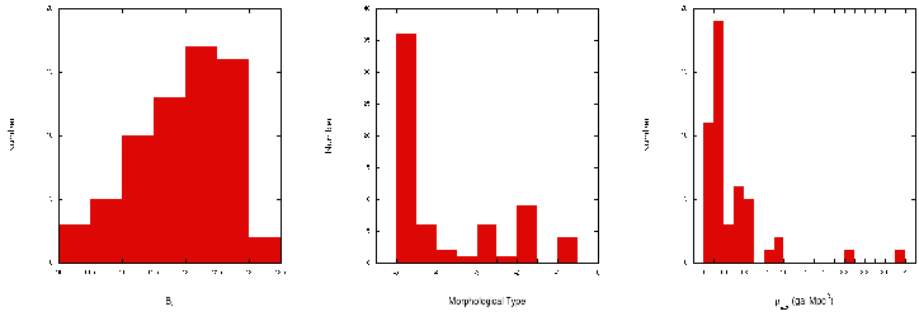}
 \caption{Distribution of B-magnitudes (left panel), morphological types 
   (middle panel), and galaxy density (right panel) for our sample  
   of 65 galaxies.}
\label{fig1} 
\end{figure}

\vspace{-0.2cm}

\section{The Sample}

The total sample of 65 galaxies includes both gas and dust--free galaxies taken from the samples of \cite{g93} and \cite{tra98}, and   
objects taken from a compilation of galaxies showing ISM traces in at 
least one of the following bands: IRAS 100 $\mu$m, X-ray, radio, 
HI and CO (\cite{ro91}).  
All galaxies belong to the {\it Revised Shapley Ames
Catalog of Bright Galaxies (RSA)} (\cite{rsa}), and
have a redshift of less than 5500 km~s$^{-1}$. 
The sample should then be biased towards objects that are expected 
to have ongoing and recent star formation, at least in small amounts, 
because of the presence of emission lines. 
On the other hand, studies of randomly selected samples of 
early-type galaxies suggest that the incidence 
of ionized-gas emission in early-type galaxies is quite high ($\sim$ 75\%, \cite{sarzi06}).
Thus, our sample should not be too dissimilar from randomly selected samples
of early-type galaxies.
Figure~\ref{fig1} summarizes the basic characteristics of the sample. 
The right panel provides evidence that a 
large fraction of galaxies is in low density environments. 
Galaxies were observed during three separate runs (March 1998, 
September 1998 and May 1999) at the 1.5m ESO telescope (La Silla).
Our long--slit spectra cover the 3700 - 7250 \AA\ wavelength range with a
spectral resolution of $\approx$7.6~\AA\ at 5550~\AA.  
For each galaxy we have extracted 7 luminosity weighted apertures (with radii: 
1.5\arcsec, 2.5\arcsec, 10\arcsec,  r$_e$/10, r$_e$/8, r$_e$/4 and r$_e$/2) 
and  4 gradients 
(0 $\leq$ r $\leq$r$_e$/16, r$_{e}$/16 $\leq$ r $\leq$r$_e$/8, 
r$_{e}$/8 $\leq$ r $\leq$r$_e$/4 and r$_{e}$/4 $\leq$ r $\leq$r$_e$/2).

\vspace{-0.2cm}


 \begin{figure*}[htp!]
 \plotone{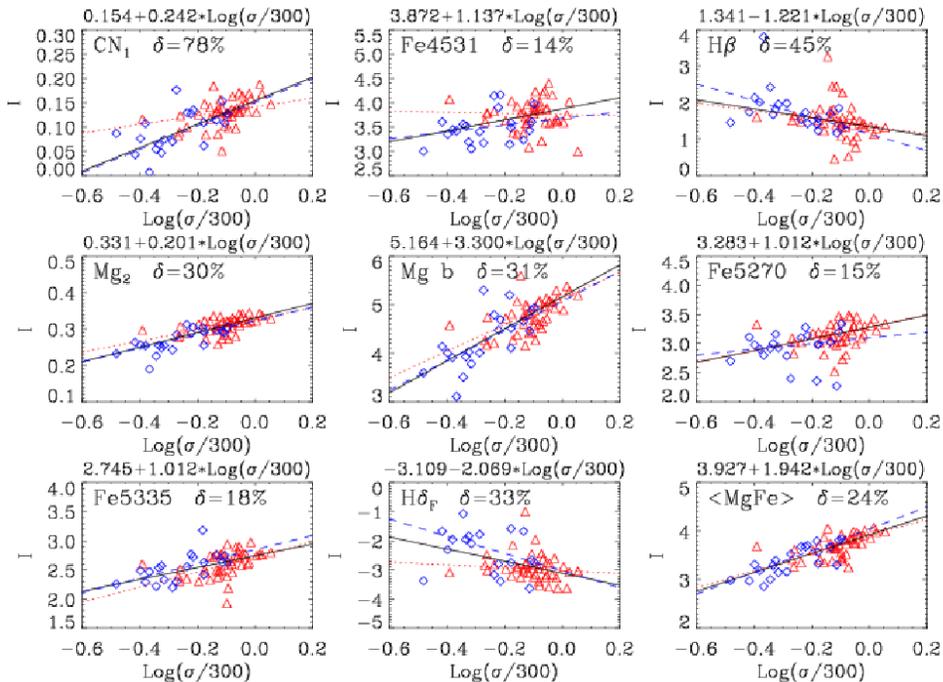}
 \caption{Selected Lick indices,  measured at $r<r_e/8$, 
for the total sample of 65 galaxies as a 
function of $\log(\sigma_c/300)$, where $\sigma_c$ is the
central velocity dispersion. Triangles and diamonds indicate E and S0 
galaxies, respectively. The dashed,
dotted, and solid lines mark the linear fit obtained for S0 galaxies, 
E galaxies, and the total sample, respectively. For each index, the linear fit to the total sample is labeled above
each panel.}
\label{fig2} 
\end{figure*}


\vspace{-0.7cm}

\section{Lick Indices}

For each aperture and 
gradient we measured 25 line--strength indices: 21 belonging to 
the original set defined by the Lick-IDS ``standard'' system 
(\cite{tra98}) and 4 subsequently introduced by 
\cite{wo97} to include the higher-order Balmer lines. 
Line--strength indices have been corrected for several effects
(contamination by possible emission, velocity dispersion) and conformed to
the Lick IDS system. 
Figure~\ref{fig2} shows selected Lick indices measured
in the central aperture ($r<r_e/8$)
for our sample as a function of central velocity dispersion.
It is clear that metallic indices 
show a well--established positive trend with 
$\sigma_c$ (with a significantly shallower variation for the  Fe indices). 
On the contrary, Balmer indices decrease with velocity dispersion.
The observed behavior suggests that metallicities or ages increase
with galaxy mass, but for the quantitative results we refer to the 
comparison with SSP models performed in Section 5.

\vspace{-0.2cm}

\section{Simple stellar population models}

To derive ages and chemical compositions from observed 
narrow band indices of galaxies, appropriate models are needed.
Thus we have derived new 
Simple Stellar Population (SSP) models for a wide range of ages
($10^9-16 \times  10^{10}$ Gyr), metallicitites 
(Z$=0.0004, 0.004, 0.008, 0.02, 0.05$) and [$\alpha$/Fe] ratios
(0--0.8). 
The SSPs are based on the Padova stellar evolution tracks
and isochrones (see \cite{bre94}). 
The standard--composition  narrow--band indices are computed on 
the basis of the
fitting functions of \cite{w94} and \cite{wo97} . 
The new $\alpha$--enhanced SSPs have been computed starting from the standard
SSPs, and accounting for the effect
of element abundance variations on the stellar atmospheres. 
In particular, the index correction due to non--solar abundance partitions 
has been derived following the main guidelines provided by previous works
in literature (\cite{tmb03}, \cite{tc04}), although 
we have revised the index dependence on  element abundance. 
We refer to Paper~III for details.
The models are available to the public at http://www.inaoep.mx/$\sim$abressan
or www.stsci.edu/$\sim$annibali/.

\vspace{-0.2cm}

\section{Ages, metallicities and [$\alpha$/Fe] ratios}
To derive stellar population parameters 
from the observed line--strength indices of our sample,
we have implemented the new $\alpha$-enhanced SSPs 
within an algorithm based on the probability density function.
The derived average ages for the whole sample, 
E and S0 are 8, 8.7, and 6.3 Gyr, respectively.
The average metallicities ([Z/H]) for the whole sample, E and S0, 
are 0.21, 0.22, and 0.19, respectively.
The average [$\alpha$/Fe] ratios for the whole sample, E and S0, 
are 0.21, 0.23, and 0.17, respectively.

\vspace{-0.2cm}

\begin{figure}
\hoffset=-10cm
\includegraphics[width=0.45\textwidth,angle=90, bb= 50 50 545 580]{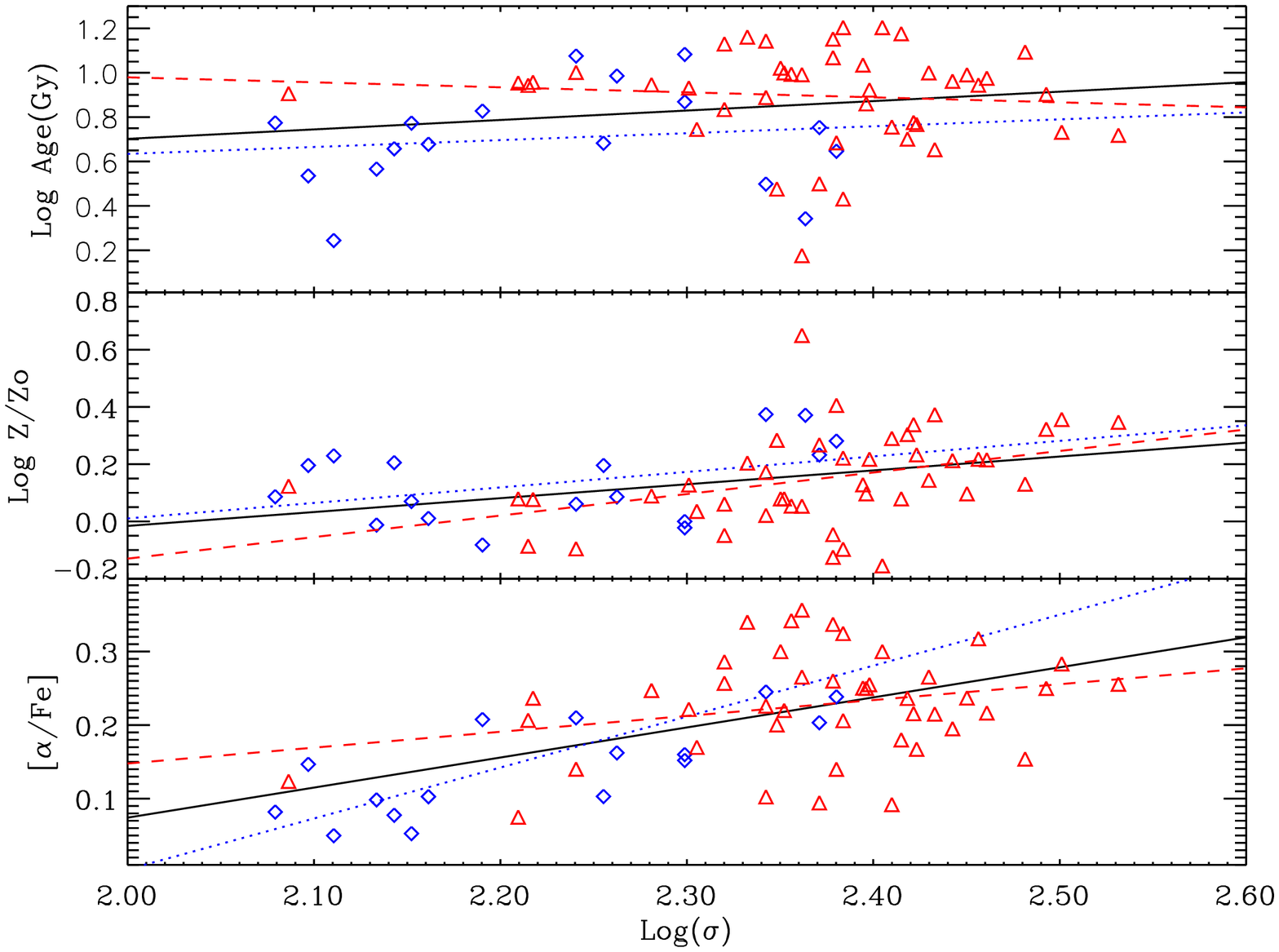}
\includegraphics[width=0.45\textwidth,angle=90, bb= 50 100 545 700]{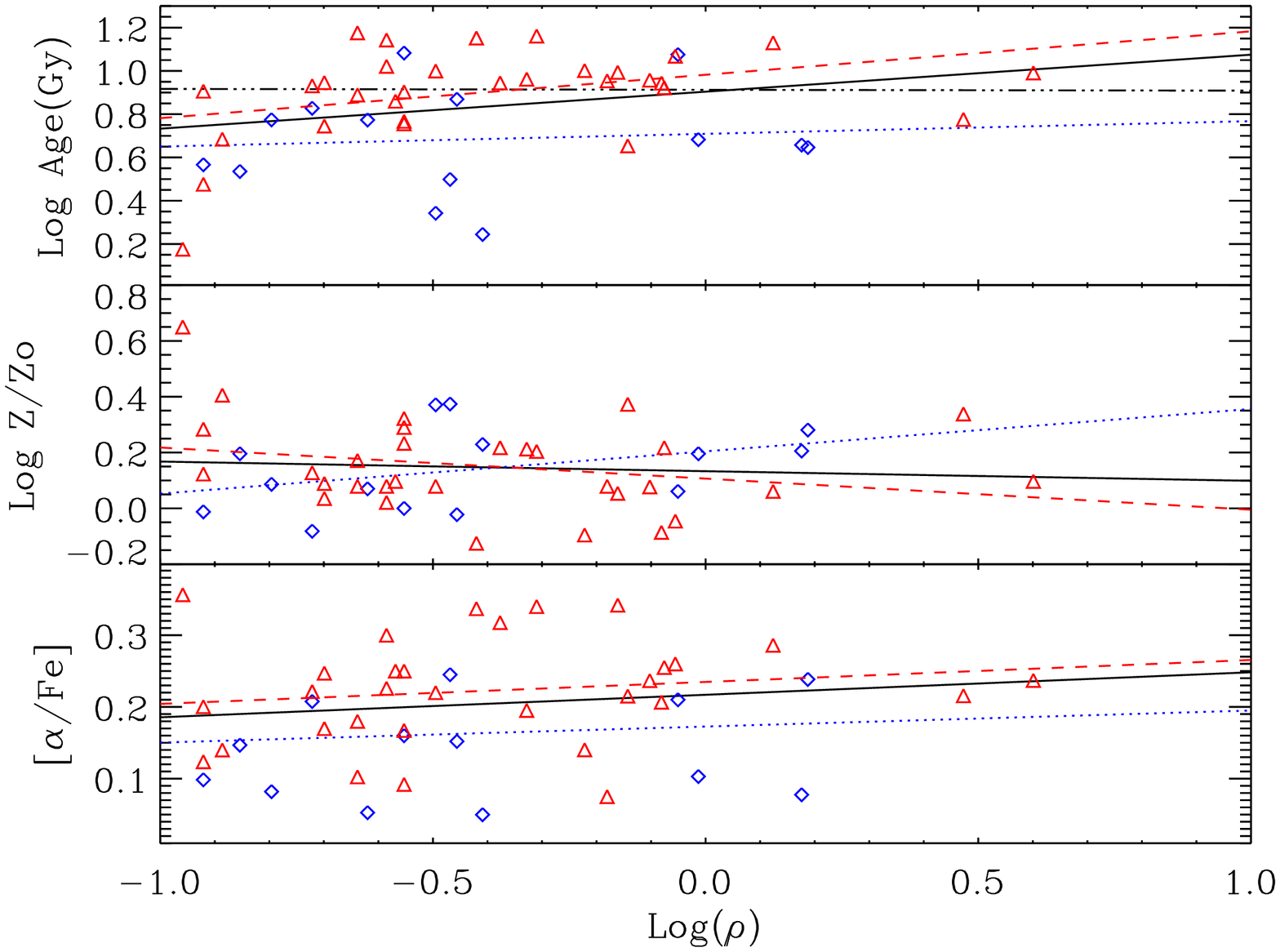}
\caption{Left panel:
 Ages, metallicities, and [$\alpha$/Fe] ratios, measured at $r_e/8$,
vs. the central velocity dispersion, $\log \sigma_c$.
Triangles and diamonds denote, respectively, E and S0 galaxies.
The solid line is the linear fit performed to all the galaxies,
while dashed and dotted lines are the best fit to Es and S0s subsamples.
Right panel: 
Ages, metallicities, and [$\alpha$/Fe] ratios, measured at $r_e/8$,
vs.  the density of the environment, $\log(\rho)$, in galaxies~Mpc$^{-3}$.}
\label{fig3} 
\end{figure}

\vspace{-0.2cm}

\subsection{Scaling Relations}

We looked for possible correlation of the stellar population 
parameters with both central velocity dispersion and local galaxy density.
In the left panel of Figure \ref{fig3} we plot ages, metallicities, 
and [$\alpha$/Fe] ratios,
derived within a $r_e/8$ aperture, as function of $\sigma_c$.
We do not find the clear signature of a global trend of age with velocity 
dispersion; on the other hand, significant positive trends are derived for 
metallicity and $\alpha$/Fe enhancement.
In the right panel of Fig.~\ref{fig3}, we plot the same  
stellar population parameters as a function of the richness parameter  
$\rho_{xyz}$, measured as galaxies $\times$ Mpc$^{-3}$.
It is interesting to note that the fit suggests a 
positive trend of age with local galaxy density. 
In particular, in spite of containing more than 
40\% of the sample with determined $\rho_{xyz}$,
the region above log($\rho_{xyz}$) $\geq$-0.4
does not contain galaxies younger than 4 Gyr.
If we consider only galaxies with ages older than 4 Gyr, 
the resulting fit is flat, indicating that
the age--environment relation is actually due to 
the presence of very young objects in the poorer environments.
This suggests that  the relations with velocity dispersion
may be "disturbed" by the presence of such 
objects populating very low density environments.
If we exclude the galaxies in very low density environments, 
we find that the relations with velocity dispersion have a
smaller scatter.

\begin{figure}
\plotone{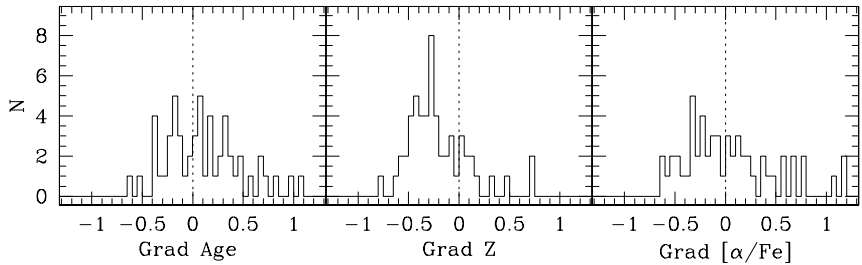}
\caption{Distributions of the age, metallicity, and [$\alpha$/Fe] gradients
computed as $\delta log(Age/Gyr)/\delta log(r/r_e)$, 
$\delta\log$(Z)/$\delta\log$(r/r$_e$), and 
$\delta$[$\alpha$/Fe]/$\delta\log$(r/r$_e$), respectively.
The dotted vertical line separates negative from positive gradient values.}
\label{fig4} 
\end{figure}

\vspace{-0.2cm}

\subsection{Gradients}

The good quality of our data allowed us also to analyze the stellar population 
gradients within the single galaxies.
The distributions of age, metallicity, and 
$\alpha$-enhancement gradients are showed in Fig.~\ref{fig4} .
While there is not clear presence of age and [$\alpha$/Fe] 
gradients, we definitely derive negative metallicity gradients (metallicity
decreases from the central regions outwards) of the order of
$\Delta \log Z/\Delta \log (r/r_e) \sim -0.21$.

\vspace{-0.4cm}

\section{Conclusions}

We have analyzed the stellar populations of a
sample of 65 early--type galaxies, mainly located in the field, 
and biased toward the presence of emission lines.
Summarizing our results:

\vspace{-0.3cm}

\begin{itemize}

\item  We derive a large age spread, from a few Gyr to 15 Gyr. 
The average SSP-equivalent ages for the whole sample, E and S0 are 8, 8.7, and 6.3 Gyr, respectively; 
the metallicity distribution shows a broad peak at $0<[Z/H]<0.3$; finally, 
the [$\alpha$/Fe] ratio definitely presents a peak at $\sim$0.22. 

\item We do not find a clear global trend of age with velocity dispersion; 
on the other hand, significant positive trends are derived for metallicity and $\alpha$/Fe enhancement.
These results testify that the chemical
enrichment is more efficient in more massive galaxies, and 
that the overall duration of the star
formation process is shorter within deeper potential wells. These two
relations do not depend on galaxy morphological type,
indicating that the galaxy gravitational potential is
the main driver of the chemical enrichment process of the galaxy.

\item Very young objects (from 1 Gyr to 4 Gyr) are found  
in very low density environments ($\rho_{xyz} \le
0.4$). None of the galaxies in high density regions is younger than 4-5 Gyr. 
The lack of environmental effect on the 
($\alpha$-enhancement)--$\sigma_c$  relation indicates that in very low
environments rejuvenation episodes, rather than more prolonged star formation,
are frequent.

\item By comparing the number of ``young'' objects with
the total number of galaxies, and by means of simple two-SSP
component models, we estimate that in these rejuvenation episodes 
(like major mergers), not more than 25\% of the galaxy mass could be formed,
that is, about 75\% of the galaxy mass 
is formed during the initial epoch of formation.

\item  We derive negative metallicity gradients within single galaxies, 
while on average the $\alpha$-enhancement remains quite flat within r$_e$/2.
This indicates that the star formation proceeded on 
typical lifetimes not significantly different across r$_e$/2, 
but evidently with a larger efficiency in the center.

\end{itemize}



\end{document}